\documentclass[
11pt,%
tightenlines,%
twoside,%
onecolumn,%
nofloats,%
nobibnotes,%
nofootinbib,%
superscriptaddress,%
noshowpacs,%
centertags]%
{revtex4-2}

\usepackage{aas_macros}
\usepackage{ljm}
\begin{document}

\titlerunning{Mathematical Modeling of K-H Instability in Partially Ionized Plasma} 

\title{Mathematical Modeling of Kelvin-Helmholtz Instability at Tangential Discontinuities in Partially Ionized Plasma: Application to Heliopause Dynamics}

\author{\firstname{A.~V.}~\surname{Titova}}
\email[E-mail: ]{avtitova@cosmos.ru} 
\affiliation{Faculty of Physics, HSE University, 20 Myasnitskaya Ulitsa, Moscow 101000, Russia}
\affiliation{Space Research Institute, Russian Academy of Sciences, Profsoyuznaya Str. 84/32, Moscow, 117335 Russia}
\affiliation{Lomonosov Moscow State University, Moscow Center for Fundamental and Applied Mathematics, GSP-1, Leninskie Gory, Moscow, 119991 Russia}
\author{{S.~D.}~\surname{Korolkov}}
\affiliation{Space Research Institute, Russian Academy of Sciences, Profsoyuznaya Str. 84/32, Moscow, 117335 Russia}
\affiliation{Lomonosov Moscow State University, Moscow Center for Fundamental and Applied Mathematics, GSP-1, Leninskie Gory, Moscow, 119991 Russia}

\author{{V.~V.}~\surname{Izmodenov}}
\affiliation{Faculty of Physics, HSE University, 20 Myasnitskaya Ulitsa, Moscow 101000, Russia}
\affiliation{Space Research Institute, Russian Academy of Sciences, Profsoyuznaya Str. 84/32, Moscow, 117335 Russia}
\affiliation{Lomonosov Moscow State University, Moscow Center for Fundamental and Applied Mathematics, GSP-1, Leninskie Gory, Moscow, 119991 Russia}

\firstcollaboration{(Submitted by G.G. Lazareva) }

\received{March 15, 2025; revised March 17, 2025; accepted April 16, 2025}

\begin{abstract} 
The stability of the heliopause, the tangential discontinuity separating the solar wind from the interstellar medium, is influenced by various processes, including the Kelvin-Helmholtz instability. This study investigates the role of charge exchange collisions between protons and hydrogen (H) atoms in reduction of the Kelvin-Helmholtz growth rate at the heliopause. Using a two-dimensional gasdynamic model with the inclusion of H atoms, we perform numerical simulations of the plasma flow near the heliopause flanks. We conduct a parametric study by varying the Knudsen number. Our results indicate that charge exchange collisions play a crucial role in suppressing the Kelvin-Helmholtz instability. As the Knudsen number decreases, the flow transitions from an unstable regime to a smoother state.

\end{abstract}



\maketitle


\section{Introduction}

While our Solar system moves through the partially ionized local interstellar medium (LISM), the solar wind -- a continuous plasma flow emanating from the Sun's upper atmosphere -- interacts with the incoming LISM flow. This interaction creates a complex structure known as the heliosheath or heliospheric boundaries, which includes the termination shock, the heliopause (a tangential discontinuity separating the solar wind from the interstellar plasma), and, presumably, a bow shock. The first two-shock model of this interaction was presented by Baranov et al. \cite{Baranov_1970}, who used the thin layer approximation to describe the fundamental structure of the heliospheric boundaries.

In the simplest gasdynamic approach, where both the stellar wind and the interstellar medium are treated as ideal gases, modeling predicts that the heliopause flanks are unstable due to Kelvin-Helmholtz instability (e.g., \cite{Korolkov_2020}), which arises from the velocity shear across the discontinuity. This instability has been widely explored using both linear analysis \cite{Baranov_1992, Ruderman_2024} and numerical modeling \cite{Korolkov_2020}. However, the physics of the heliosheath is more complicated, and additional physical processes may influence the stability of the heliopause. For example, studies have shown that the external interstellar magnetic field can suppress the growth of Kelvin-Helmholtz instability and contribute to flow stabilization \cite{Decin_2013, Meyer_2021, Florinski_2005}. Furthermore, the periodic nature of the solar wind also plays a stabilizing role in reducing the Kelvin-Helmholtz instability \cite{Korolkov_2022}.

As was mentioned above, the LISM is partially ionized with its neutral component mainly consisting of hydrogen (H) atoms. These atoms interact with protons through charge exchange and significantly affect the flow pattern of the heliosheath, especially the flanks \cite{Aleksashov_2004}. So, they may have an influence on the stability of the heliopause as well. Chalov \cite{Chalov_2018, Chalov_2019} investigates the linear instability of a tangential discontinuity in a partially ionized hydrogen plasma using the linear analysis. The study assumes a plasma layer divided into two regions separated by a tangential discontinuity, with hydrogen atoms being collisionless and interacting with protons through charge exchange. The lower plasma is in thermodynamic equilibrium with the hydrogen atoms, while the upper plasma is assumed to have no charge exchange. The study identifies several instability modes: central modes, flank modes, and classical Kelvin-Helmholtz modes. Central modes are new unstable modes that appear due to charge exchange and do not explicitly depend on the relative velocity of the two plasmas. This instability is called dissipative instability. The charge exchange process modifies the classical Kelvin-Helmholtz instability, reducing the instability increment. However, the dissipative instability occurs when the dissipative processes are taken into account. Therefore, contrary to our intuition, dissipation should destabilize tangential discontinuities. 

Another study addressing the influence of charge exchange was performed by Avinash et al. \cite{Avinash_2014}. The authors develop a general model that includes all essential physical processes, such as charge exchange and magnetic fields to analyze the stability of the heliopause. The study derives a general dispersion relation that is used to analyze the stability of different regions of the heliopause, including the nose, shoulder, and flanks. The results show that the nose region is unstable to Rayleigh-Taylor-like modes, while the shoulder region shows a mixed instability. On the flanks, the Kelvin-Helmholtz-like instability is stabilized by magnetic fields but is re-activated by the presence of energetic neutral atoms (ENAs), which are generated by charge exchange interactions in the heliosheath. The same effect was observed by Borovikov et al. \cite{Borovikov_2008}.

\begin{figure}[t]
    \centering
    \includegraphics[width=1\textwidth]{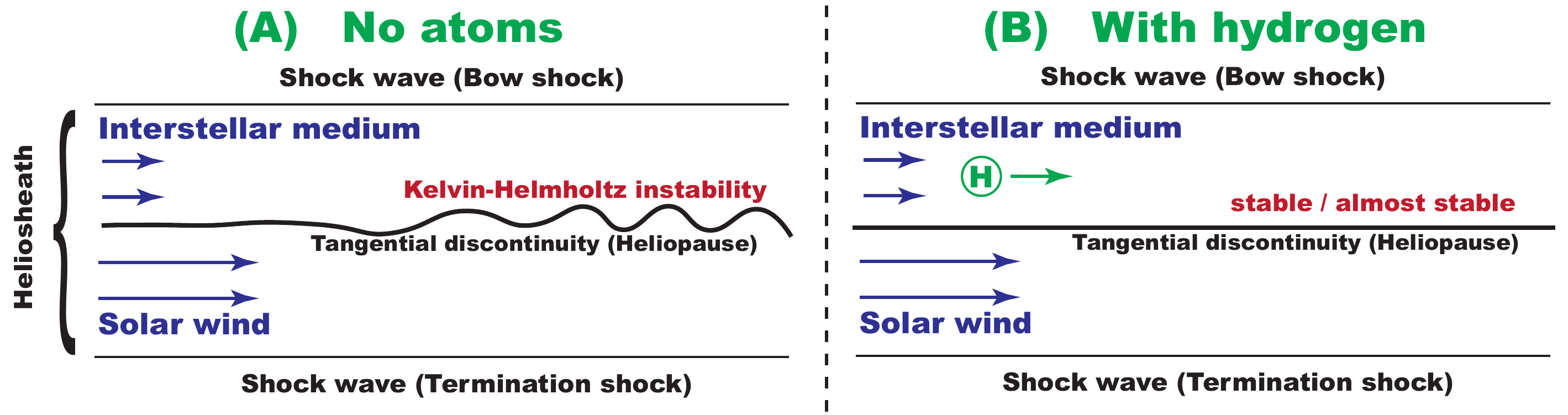}
    \caption{Schematic picture of the problem}
    \label{fig:scheme}
\end{figure}

The main aim of this study is to investigate numerically the effects of charge exchange on the Kelvin-Helmholtz instability of the heliopause flanks. To achieve this, we conduct a parametric study focusing on a simplified model of two parallel plasma flows separated by a tangential discontinuity (see Figure \ref{fig:scheme}). We employ a purely gasdynamic approach, neglecting magnetic fields and solar wind periodicity, to isolate the influence of charge exchange from other stabilizing effects.

The rest of the paper is organized as follows: Section \ref{Model} describes the mathematical formulation of the problem. In Section \ref{num_approach}, we discuss the computational method. The results of our simulations are presented in Section \ref{results}, followed by conclusions in Section \ref{conclusions}.
\section{Model}\label{Model}

To explore the Kelvin-Helmholtz instability, we propose a simplified model with two parallel plasma flows separated by a tangential discontinuity in a channel with width $L$, analogous to the inner and outer heliosheath regions bounded by the termination and bow shocks (Figure \ref{fig:scheme}). The plasma, consisted of electrons and protons, is assumed to be the ideal non-heat-conducting gas with constant heat capacity, adiabatic index $\gamma=5/3$, and is described by Euler equations. A single fluid approach is used with the equation of state: $p = 2n_pk_bT$, where $p$ is the pressure, $n_p$ is the plasma number density, $k_b$ is the Boltzmann constant, and $T$ is the temperature. Additionally, the system includes a neutral hydrogen component that interacts with the plasma through charge exchange.

The problem is set in a two-dimensional spatial domain. At the inlet boundary $x=0$ we fix the values according to the boundary conditions, distinguishing between two layers: the upper layer ($y>L/2$) represents the outer heliosheath with uniform velocity $\mathbf{v_1}=v_1\cdot\mathbf{e_x}$ , number density $n_1$, and pressure $p_1$. The lower layer ($y<L/2$) corresponds to the inner heliosheath, characterized by a higher velocity  $\mathbf{v_2}=v_2\cdot\mathbf{e_x}$, lower number density $n_2$, while maintaining the same pressure $p_2=p_1=p$ as the upper layer. The parameters at the inlet boundary are adapted from the global model of the heliosphere \cite{Izmodenov_2020}: $v_1=26.4$ km/s, $n_1 =0.013$ cm$^{-3}$, $p=2.3\cdot10^{-13}$ Ba, $v_2=66$ km/s, $n_2 =0.002$ cm$^{-3}$, and the width of the channel is assumed to be $L=400$ AU.   At the channel outlet, soft boundary conditions are applied. On the upper and lower boundaries, the pressure $p$ is fixed, while soft conditions are imposed for the velocity and density.

In this paper, we neglect the influence of the charge exchange process on the dynamics of the H atoms, as the primary goal is to study the influence on the plasma flow. For that reason, the parameters of the H atom flow -- number density $n_H$, pressure $p_H$, and velocity $\mathbf{V_H}=V_H\cdot\mathbf{e_x}$ -- are assumed to be constant.

The corresponding Euler equations can be written as:

\begin{equation}
\begin{cases} 
 \frac{\partial \rho}{\partial t} +\text{div}(\rho \mathbf{V}) = 0, \\ 
  \frac{\partial (\rho \mathbf{V})}{\partial t}+ \text{div}(\rho \mathbf{VV} + p \hat{\mathbf{I}}) = \mathbf{Q}_2, \\ 
   \frac{\partial E}{\partial t}+\text{div}((E + p) \mathbf{V}) = {Q}_{3}.
\end{cases}
\end{equation}

where $\rho=m_pn_p$, $\mathbf{V}$, \( E = \frac{p}{\gamma - 1} + \frac{\rho V^2}{2} \) are the plasma density, velocity, and the energy density correspondingly. The expressions for source terms are written in the form of McNutt, Lyon \& Goodrich \cite{McNutt_1998}:

\[
    \mathbf{Q}_2 = \nu_H \cdot (\mathbf{V}_H - \mathbf{V}), \quad {Q}_{3} = \nu_H \cdot \left[ \frac{U_H^*}{U_{M,H}^*} \left(c_H^2 - c^2\right) +\frac{{V^2_H} - {V^2}}{2} \right], \text{ where}
    \]

\[
    \nu_H = \frac{1}{m_p} \rho \rho_H U_{M,H}^* \sigma_\text{ex}(U^*, M, H), \quad
    c = \sqrt{\frac{2k_B T}{m_p}}, \quad c_H = \sqrt{\frac{2k_B T_H}{m_p}},
\]
\[
    U_H^* = \sqrt{|\mathbf{V_H} - \mathbf{V}|^2 + \frac{4}{\pi} \left( c_H^2 + c^2 \right)}, \quad 
    U_{M,H}^* = \sqrt{|\mathbf{V_H} - \mathbf{V}|^2 + \frac{64}{9\pi} \left( c_H^2 + c^2 \right)}.
\]

Here, $\rho_H$, $T_H$, $\mathbf{V_H}$ are the number density, temperature, and bulk velocity of H atoms, and $\sigma_\text{ex}$ is the total charge exchange cross section. In this paper, we take the cross section by Lindsay and Stebbings (2005), which can be written as $\sigma_{ex}(U)=(a_1-a_2\text{ ln } U)^2$ ($U$ is the relative speed in cm/s), where $a_1=2.2835\cdot10^{-7}$, $a_2=1.062\cdot10^{-8}$.

To prevent plasma deceleration over distance, the following conditions are imposed for H atoms:
\[
{V}_H={v}_1, \quad {T}_H={T}_1\quad {n}_H=\alpha\cdot{n_1},
\]

where $\alpha$ is a parameter showing the ratio between the H atom number density and the proton number density.

We also adopt the same assumption as Chalov \cite{Chalov_2018}, considering charge exchange only in the outer heliosheath (upper) layer. This is mathematically equivalent to setting the charge-exchange cross section in the lower layer to zero ($\sigma_{ex2}=0$). This approximation is physically justified when the plasma in the lower layer is much hotter. In our case, $T_2/T_1=\rho_1/\rho_2\approx6.7$.

The problem depends on the following dimensional parameters: $L$, $\rho_1$, $p$, $v_1$, $\rho_2$, $v_2$, $\alpha\cdot\sigma_{ex1}(v_1)$. To cast the problem in a dimensionless form, we relate the density to $\rho_1$, the pressure to $p$, and the distances to the width of the channel $L$. In that case, the boundary conditions are transformed to $\hat{\rho}_1=\hat{p}=1, \hat{v}_1=M_1\sqrt{\gamma}$ in the upper layer and $\hat{\rho}_2=\rho_2/\rho_1=\eta$, $\hat{p}=1$, $\hat{v}_2=M_2\sqrt{\gamma/\eta}$ in the lower layer. $M_1$ and $M_2$ are the Mach numbers in the upper and lower layers. 

The resulting dimensionless system of equations is given by:
\begin{equation}
\begin{cases}
    \frac{\partial \hat{\rho}}{\partial \hat{t}}+\text{div}(\hat{\rho} \mathbf{\hat{V}}) = 0, \\
     \frac{\partial (\hat{\rho} \mathbf{\hat{V}})}{\partial \hat{t}}+\text{div}(\hat{\rho} \mathbf{\hat{V}} \mathbf{\hat{V}} + \hat{p} \mathbf{\hat{I}}) = \dfrac{1}{\text{Kn}} \, \mathbf{\hat{Q}_2}, \\
     \frac{\partial \hat{E}}{\partial \hat{t}}+\text{div}((\hat{E} + \hat{p}) \mathbf{\hat{V}}) = \dfrac{1}{\text{Kn}} \hat{Q}_{3},
\end{cases}
\begin{cases}
    \mathbf{\hat{Q}_2} = \hat{\nu}_H \cdot \left( \mathbf{\hat{V}_H} - \mathbf{\hat{V}} \right), \\
    \hat{Q}_{3} = \hat{\nu}_H \left[ \dfrac{\hat{U}_H^{\ast}}{\hat{U}_{M,H}^{\ast}} (\hat{c}_H^2 - \hat{c}^2) \right], \\
    \hat{\nu}_H = \hat{\rho} \hat{U}_{M,H}^{\ast} \sigma_{ex}(\hat{U}_{M,H}^{\ast}), \quad \sigma_{ex}(\hat{U}) = \left(1 - \hat{a}_2 \cdot \ln(\hat{U}) \right)^2, \\
  \quad \hat{a}_2 = \dfrac{a_2}{\sqrt{\sigma_{ex}(\hat{U})}}
\end{cases}
\end{equation}

where $\text{Kn}=l/L$ is the Kundsen number that represents the ratio of $l = 1/(\alpha n_1 \sigma_{ex})=1/(n_H\sigma_{ex}(v_1))$, which is the mean-free path of protons before charge exchange with atoms, to the characteristic size $L$ of the problem. The regime with no atoms, as seen from the equations, is achieved at $\text{Kn}\rightarrow\infty$.

As a result, the problem depends on the following dimensionless parameters:
\[
\eta \quad M_1, \quad M_2, \quad \text{Kn}.
\]

From the dimensional values on the boundary we get:
\[
\eta=0.15, \quad M_1=0.63, \quad M_2\approx0.97M_1\approx0.61.
\]

The Kn parameter will be varied by different values.

\section{NUMERICAL SIMULATION } \label{num_approach}
In this study, we adopt the computational methodology outlined in \cite{Korolkov_2020, Titova_2021}. Our approach employs a finite volume Godunov-like scheme.

The simulations are conducted in a two-dimensional Cartesian coordinate system, where the computational domain is divided into a grid of cells. For each cell, we solve the governing equations of mass, momentum, and energy conservation. To accurately compute the fluxes of mass, momentum, and energy across the cell interfaces, we use the HLLC (Harten-Lax-van Leer-Contact) Riemann solver \cite{Miyoshi_2005}. To enhance the accuracy of the numerical scheme, we implement the minmod limiter (a Total Variation Diminishing (TVD) scheme \cite{Hirsch_1990}).

The computational domain is resolved with a grid resolution of 
$4096\times1024$ cells. This resolution is achieved by using CUDA technology of parallel programming on video cards. Nevertheless, alternative resolutions were also checked. The influence of the numerical solution on both the chosen method and grid resolution was investigated in \cite{Korolkov_2020}

\section{SIMULATION RESULTS}\label{results}
This section presents the outcomes of our numerical simulations. We focus on the parameter $\text{Kn}$ and vary it to study its influence on the instability. Firstly, we will present the 2D distributions of proton parameters, then we will analyze the Fourier Transform at chosen points to better characterize the fluctuations obtained in different solutions.
\begin{figure}[t]
    \centering
    \includegraphics[width=1\textwidth]{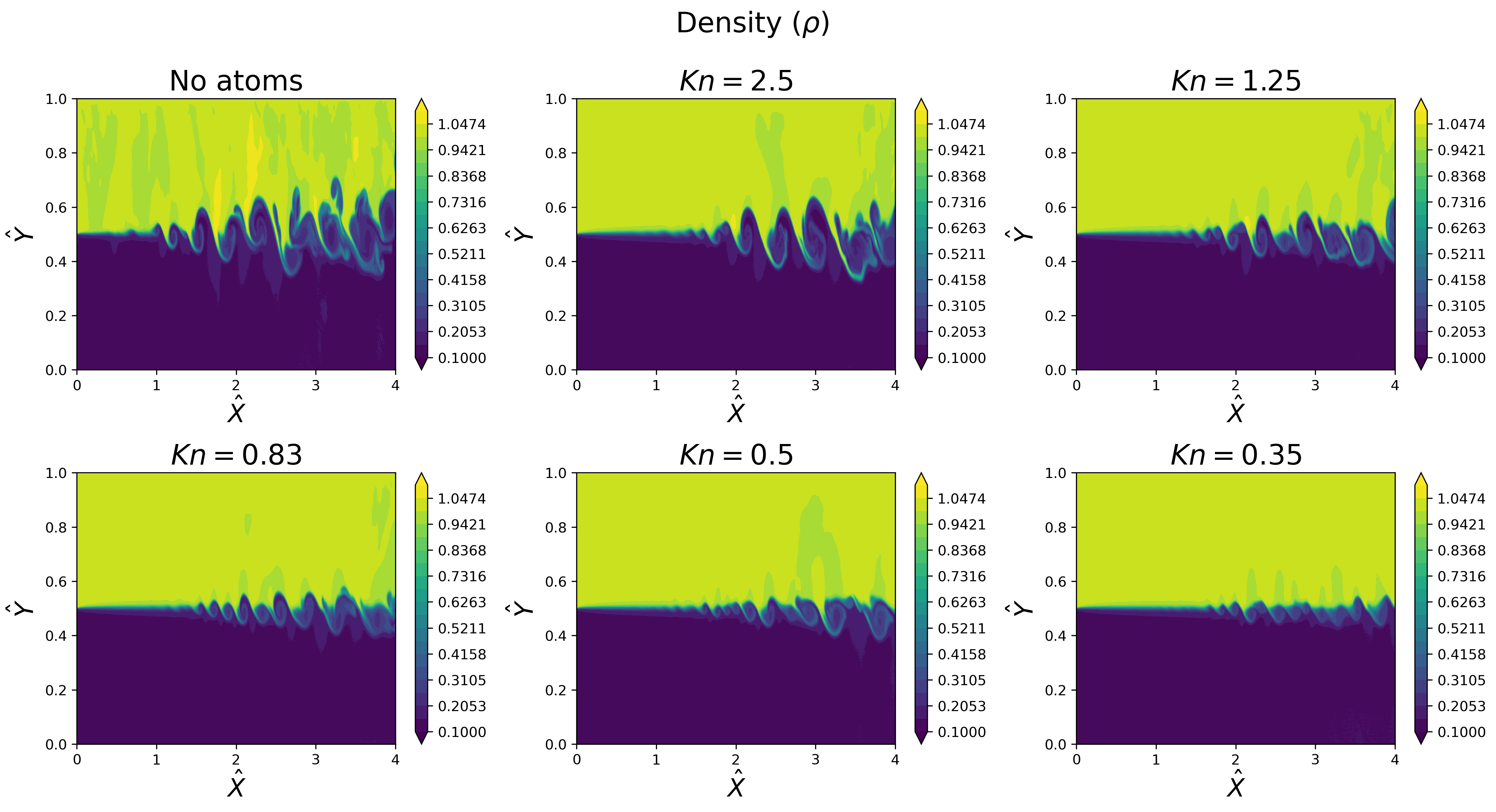}
    \caption{Plasma density distribution (\(\hat{\rho}\)) for different \(\text{Kn}\) numbers. }
    \label{fig:1}
\end{figure}
\begin{figure}[t]
    \centering
    \includegraphics[width=1\textwidth]{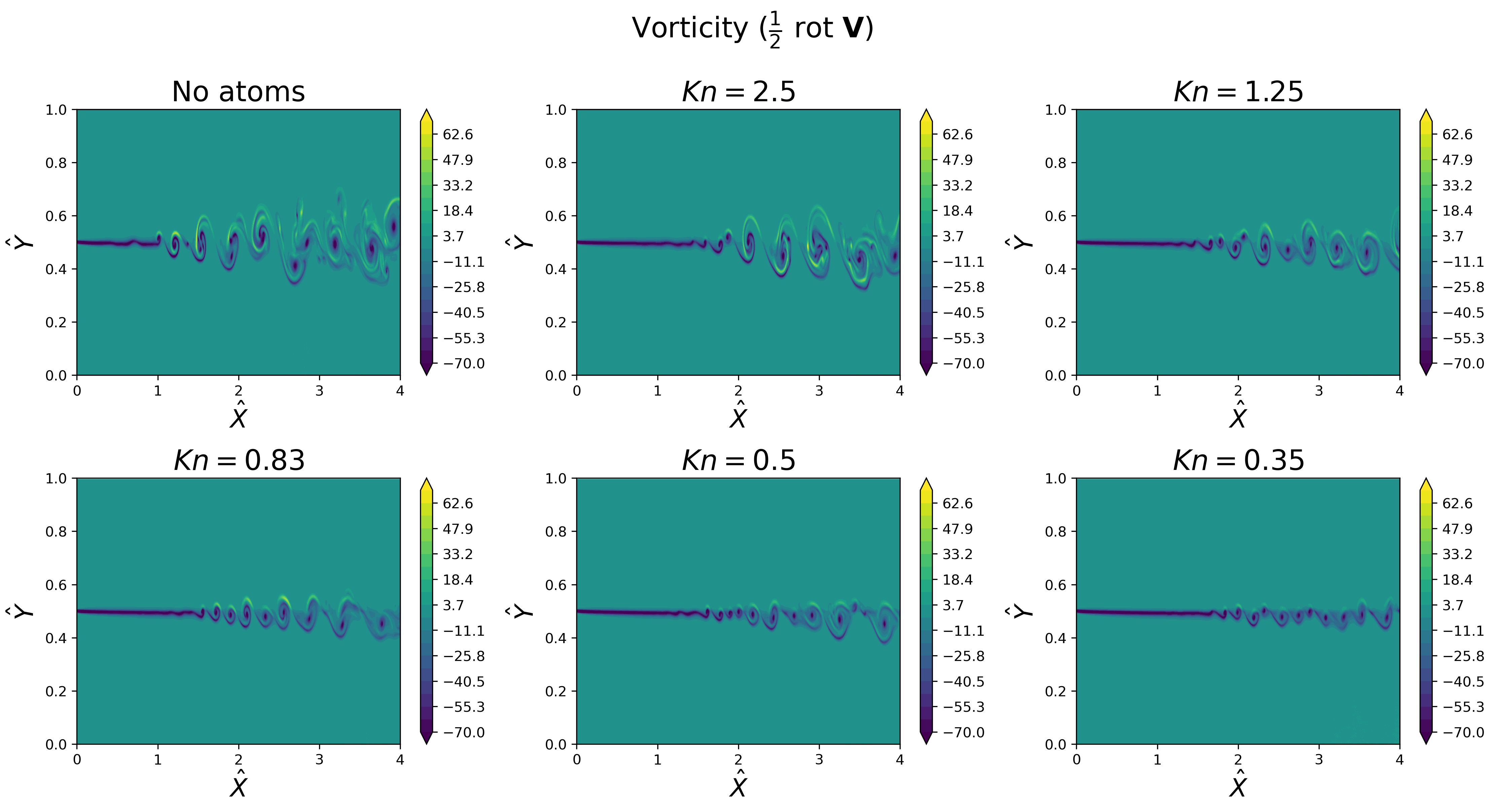}
    \caption{The vorticity (\(\omega=\frac{1}{2} \text{rot} \mathbf{\hat{V}}\)) isolines for different \(\text{Kn}\) numbers. }
    \label{fig:2}
\end{figure}

Figures \ref{fig:1} and \ref{fig:2} presents the plasma density distribution $\hat{\rho}$ and vorticity isolines for different $\text{Kn}$ numbers. The heliopause structure is prominently visible at $\hat{y} = 0.5$.

In the absence of H atoms ($\text{Kn}\rightarrow\infty$), the plasma flow is significantly unstable starting from $\hat{x} \approx 0.6$. This instability manifests as large scale vortices, which are typical for the Kelvin-Helmholtz instability. The Kelvin-Helmholtz instability arises due to shear flow between plasma regions with different velocities, leading to the formation of vortices and waves. These disturbances are clearly visible in the density distribution. 

At $\text{Kn} = 2.5$, disturbances decrease significantly, and vortices become less pronounced. As $\text{Kn}$ decreases, the influence of the energy sources and momentum increases. The stabilizing effect of charge exchange collisions becomes evident. At $\text{Kn}=0.5$, the flow appears significantly smoother, with fewer visible disturbances. By $\text{Kn} = 0.35$, the instability, while still present, is reduced to very low levels.

So, the charge exchange collisions reduce the Kelvin-Helmholtz instability as predicted by the linear analysis by Chalov \cite{Chalov_2018}, however, in our calculations the flow is stabilized and no other types of instabilities occur. 

We should also note that for very low Kn numbers (e.g. Kn = 0.35 and lower), the influence of H atoms is very strong; therefore, for such cases, this simplified approach may not be valid, as the H atom parameters cannot be considered constant. This could potentially affect the stability of the Kelvin-Helmholtz instability.

\begin{figure}[t]
    \centering
    \includegraphics[width=1\textwidth]{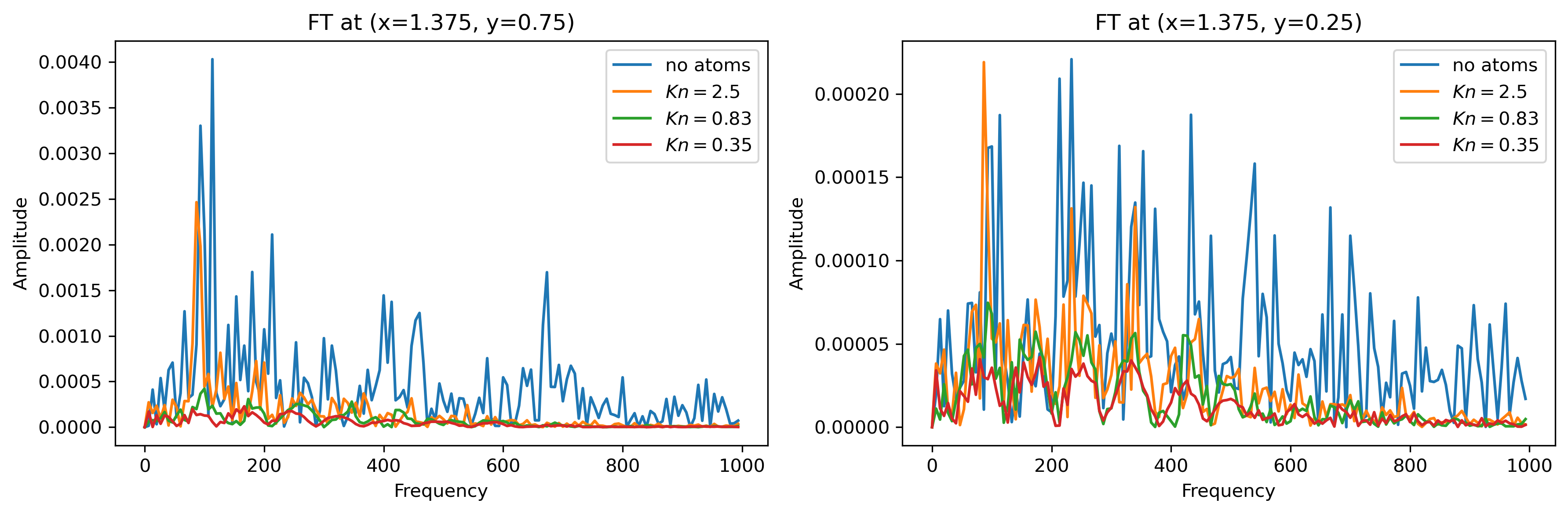}
    \caption{Fourier transform at $\hat{x}=1.375, \hat{y}=0.75$ (left panel) and $\hat{x}=1.375, \hat{y}=0.25$ (right panel) for various $\text{Kn}$ numbers}
    \label{fig:3}
\end{figure}

Figure \ref{fig:3} shows the Fourier transform of plasma density fluctuations over time at two spatial locations -- Point A ($\hat{x}=1.375, \hat{y}=0.75$) in the outer heliosheath (interstellar medium side) and Point B ($\hat{x}=1.375, \hat{y}=0.25$) in the inner heliosheath (solar wind side) -- across varying $\text{Kn}$ values.

At Point A, the Fourier spectrum reveals significant harmonic activity when $\text{Kn}=0$, with dominant amplitudes at low frequencies. These low-frequency peaks signal large-scale instabilities or oscillations driven by velocity shear at the heliopause boundary. As $\text{Kn}$ decreases to 2.5, the harmonic amplitudes diminish, reflecting partial stabilization through charge exchange collisions. By $\text{Kn}=0.83$, the peaks nearly vanish, demonstrating near-complete suppression of density fluctuations. At $\text{Kn}=0.35$ no high peaks are observed, which means that under the influence of charge exchange, complete damping occurred. A similar trend is observed at Point B, where decreasing 
$\text{Kn}$ similarly reduces harmonic amplitudes. 

\begin{figure}[t]
    \centering
    \includegraphics[width=1\textwidth]{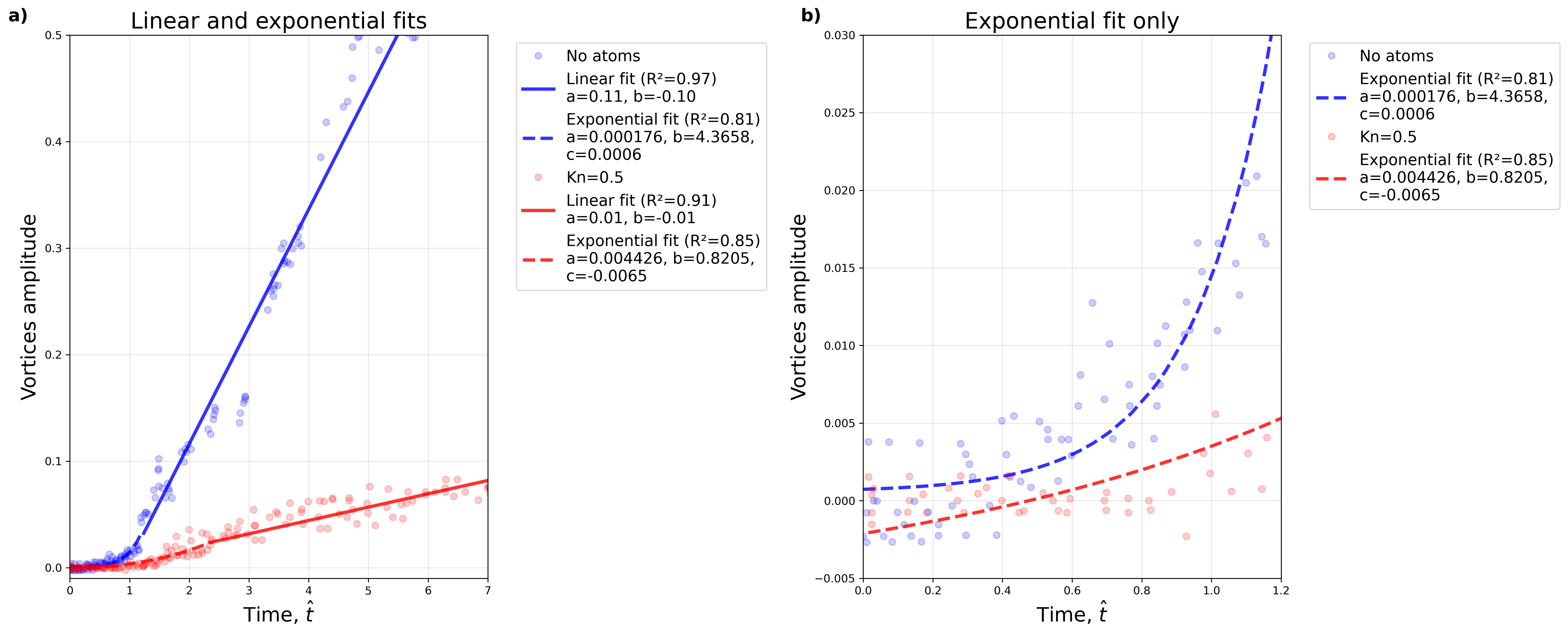}
    \caption{ Time evolution of vortex amplitude for two cases: without atoms (blue) and a case with Knudsen number Kn=0.5 (red).
(a) Both linear (solid lines) and exponential (dashed lines) fits are applied to the data. (b) Zoomed in figure at small times with exponential fits only. Fitted parameters are provided in the legend.
}
    \label{fig:4}
\end{figure}

Figure~\ref{fig:4} illustrates the evolution of large-scale vortices amplitude as a function of non-dimensional time \( \hat{t} \), calculated by dividing the spatial coordinate \( \hat{x} \) by the average velocity of the large-scale vortices (\( \hat{V}_{\text{vortices}} \approx 1.42 \)). To obtain a greater number of data points, values were extracted at several time points during which the flow remained quasi-stationary. The figure compares two cases: a gas dynamic case without atoms and a kinetic case with Kn= 0.5. In both cases, the vortex amplitude increases with distance; however, the gas dynamic case shows a significantly more pronounced growth, while the Kn=0.5 case demonstrates a more modest increase in amplitude.

According to linear stability theory, the growth of perturbations across a tangential discontinuity is expected to follow an exponential law (e.g.,~\cite{Kulikov_2019}) . The characteristic time for the development of the Kelvin--Helmholtz instability is given by:
\[
\tau_{\text{KH}} = \frac{(\rho_{1} + \rho_{2})}{v_{\text{rel}} k \sqrt{\rho_{1} \rho_{2}}},
\]
where \( k \) is the wavenumber and \( v_{\text{rel}} \) is the relative velocity of the two layers. For our gas dynamic case, this expression yields a typical instability growth time of \( \tau_{\text{KH}} \approx 0.12 \) in dimensionless units, corresponding to approximately 8.6 years.

To investigate the large-scale vortices development behavior in our simulations, we perform curve fitting using two functions: a linear model \( f = a t + b \) and an exponential model \( f = a \exp(bt) + c \). The fitted curves and corresponding coefficients are shown in Figure~\ref{fig:4}. In the early-time stage (Figure~\ref{fig:4}b), the amplitude grows exponentially. However, as the vortices become large and fully developed, linear theory no longer applies, and the growth becomes linear. 

The early-time exponential fit enables us to estimate the instability growth time as \( \tau_{\text{KH}} \approx 1/b \). For the gas dynamic case, this yields \( \tau_{\text{KH}} \approx 0.23 \) (16.6 years), which is approximately twice the value predicted by linear theory. In the case with \( \mathrm{Kn} = 0.5 \), the estimated growth time increases significantly to \( \tau_{\text{KH}} = 1.22 \) (87.8 years).

\section{Conclusions} \label{conclusions}

This study has explored the effects of charge exchange collisions on the Kelvin-Helmholtz  instability at the heliopause using a simplified two-dimensional gasdynamic model. By varying the Knudsen number, we have demonstrated that charge exchange plays a crucial role in stabilizing the plasma flow. In the absence of hydrogen atoms (\(\text{Kn} \rightarrow \infty\)), the heliopause shows significant instability, characterized by pronounced vortices driven by velocity shear. As Knudsen number decreases, the influence of the sources rises. This process suppresses the instability and stabilizes the flow. 

The stabilization mechanism is further supported by the analysis of Fourier transforms of density fluctuations, which show a clear transition from high-amplitude oscillations at high Knudsen numbers to almost uniform, low-amplitude profiles at low Knudsen numbers. 

\section*{Acknowledgements}

%

\end{document}